# Performance Peculiarities of Viterbi Decoder in Mathworks Simulink, GNU Radio and Other Systems with Likewise Implementation

Alexey Shapin, Denis Kleyko, Nikita Lyamin, Evgeny Osipov and Oleg Melentyev

*Abstract*—The performance of convolutional codes decoding by the Viterbi algorithm should not depend on the particular distribution of zeros and ones in the input messages, as they are linear. However, it was identified that specific implementations of Add-Compare-Select unit for the Viterbi Algorithm demonstrate the decoding performance that depends on proportion of elements in the input message. It is conjectured that the modern commercial hard- and software defined communication equipment may also feature similar implementation and as such their decoding performance could also vary.

*Index Terms*— Add-Compare-Select, Viterbi algorithm, convolutional codes

## I. INTRODUCTION

CONVOLUTIONAL codes are a class of error-correcting codes. They are widely used in modern telecommunication systems, e.g. in IEEE 802.11 [1], IEEE 802.16 [2], UMTS [3], LTE [4], etc. Convolutional codes can be decoded by several methods, e.g. Viterbi algorithm, sequential decoding, BCJR algorithm. The Viterbi algorithm (VA) [5] is one of the most common approaches, it is recommended in all the above mentioned standards. It is also implemented in Mathworks Simulink Communications Systems Toolbox [6] and GNU Radio [7] which are widely used for experiments and a performance evaluation of digital communication systems by the research community.

There are many studies of different aspects of the VA, however, this paper only investigates how specifics of implementation of the VA affect the decoding performance.

The convolutional codes are linear codes and theoretically bits of the input message prior encoding should not affect the performance of the VA. However, throughout this paper it is shown that in specific implementations of the VA the proportion between zeros and ones in the message may influence the decoding performance. Thus, this letter reports that under various implementations of Add-Compare-Select unit there may be dependency between the performance of the Viterbi decoder and the proportion of bits in the input message. The presented finding is by itself useful to a wide community of researchers and practitioners while assessing the performance of communication systems via simulations.

The contribution of this paper is threefold:
- it was identified that specific implementations of Add-Compare-Select unit for VA demonstrate the decoding performance that depends on proportion of zeros and ones in input messages;
- the theoretical influence of the discovered effect on the decoding performance of various Viterbi decoder implementations was studied;
- it is conjectured based on available literature that the decoding performance of the modern decoding hardware could also vary due to specifics of the VA implementation.

The letter is structured as follows. The implementation details of the Viterbi algorithm are described in Section II. The problem formulation follows in Section III. Section IV presents the simulations methodology and the results. The letter is concluded with a discussion on hardware decoders and potential for performance improvements in Section V.

## II. RELEVANT IMPLEMENTATION DETAILS OF VITERBI ALGORITHM

For the detailed description of the Viterbi algorithm as well as its extensive performance analysis in the context of decoding of convolutional codes over a noisy communication channel the interested reader is referred to [5, 8]. The VA was firstly identified as the maximum likelihood sequence estimator in [9]. This section recalls the major aspects of the algorithm, which are relevant for the considered problem. In general the VA finds the most likely message of states of a Hidden Markov Model, called a *path*, which produces a particular observed output. Each decoding step the VA produces several candidate *branching paths,* which lead to the same output state. This is illustrated in in Fig. 1. Each branch is characterized by a *branch metric*, which when summed up result in a *path metric*. The algorithm selects the path with the *minimal* path metric. Thus, the VA operations are implemented in the Add-Compare-Select (ACS) unit [9] illustrated in Fig. 2.

The comparator produces three different outcomes: "less", "greater" or "equal". While the decision in the first two cases is univocal, handling of the "equal" case differs from one implementation to another resulting in a rather different performance. The decision-making in the case of equality of the path metrics is the major focus of this letter. Specifically, it is demonstrated that this situation appears often when hard-input decoder or small number of quantization levels for the soft-input decoder are considered.

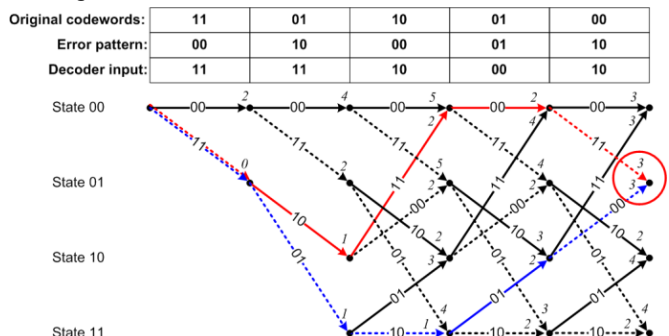

Fig. 1. Paths' metrics equality in the considered example.

Fig. 1 demonstrates an example of the metric equality case occurrence on the trellis for simple convolutional code with the constraint length 2 and polynomials in octal form (5, 7). The following notations were adopted: branching paths caused by a *zero* input bit are shown as solid lines and branching paths caused by a *one* input bit are shown as dashed lines, similarly with [5]. At the end of the trellis in the state "01" denote the top path "10001" as p1 (marked with red), the bottom path "11101" as p2 (marked with blue) and paths metrics as M1 and M2 correspondingly. Note that p1 has more zeros compare to p2; M1 and M2 are equal.

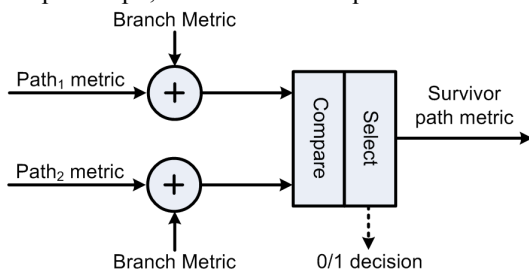

Fig. 2. A schematic of the ACS operation.

### III. PROBLEM FORMULATION

According to the original VA, in the metric equality case the surviving path must be selected with equal probability by tossing a coin [5]. This property is of the major importance and makes the decoding performance of VA independent from the proportion of the bits in the input message. In the example demonstrated in Fig. 1 according to the original VA both paths have equal chances to survive. However, many VA implementations including Simulink [6], GNU Radio [7] use strict rule instead of random path selection as supposed by the original VA. Available literature on modern hardware decoders [10,11,12,13,14,15,16] gives us evidence that at least part of them also uses strict rule.

There are two options for the strict path selection implementation: the zero-oriented and the one-oriented approaches.

The *zero-oriented* approach is based on the "strictly greater" condition. If M1 > M2, then p2 is selected, otherwise (including the equality case) p1 is selected. In the example on the trellis (Fig. 2) it can be shown that such Viterbi decoder selects the path, which is closer to the top (p1). Hence, the path containing more zeros is chosen. This pre-defined deterministic choice makes the VA to be zero-oriented. This letter demonstrates that for such implementation the higher the proportion of zeros in an input message the lower is the bit error rate (BER) after decoding under the same Signal-to-Noise ratio (SNR) in the channel.

The *one-oriented* approach is based on the "strictly less" condition. If M1 < M2, then p1 is selected, otherwise (including the metric equality case) p2 is selected. In the discussed example such Viterbi decoder selects the path, which is closer to the bottom of the trellis (p2) or in other words, the path with more ones. This letter demonstrates that for such implementation the higher the proportion of ones in an input message the lower is BER after decoding for the same SNR in the channel.

The intuition behind this dependency is that the path coming from the upper part of the trellis contains more zeros than the path coming from the lower part of the trellis. Note that despite the fact that the last bits of the paths leading to the same state on the trellis are the same (e.g. in Fig. 2 both p1 and p2 are terminated by one), the choice of the particular path will affect the preceding bits. In other words the effect can be seen in retrospective only. Thus, if a path with more zeros survives and zeros dominate in the input message then some bits will be correctly "guessed" by the decoder; similarly, if a path with more ones survives and zeros dominate in the input message then some bits will be wrongly "guessed" by the decoder.

### IV. SIMULATION METHODOLOGY AND RESULTS

In this section we demonstrate the theoretical influence of the discovered effect on the decoding performance of various Viterbi decoder implementations through a series of simulations performed using Simulink. In order to study the influence of the proportion of the input bits on the performance of different Viterbi decoders, Binary Bernoulli Generator (BBG) with controlled input probability of zero (PoZ) was used. Random paths selection and the ones-oriented Viterbi decoders were implemented as 2-level S-functions. The Viterbi decoder from the Communications System

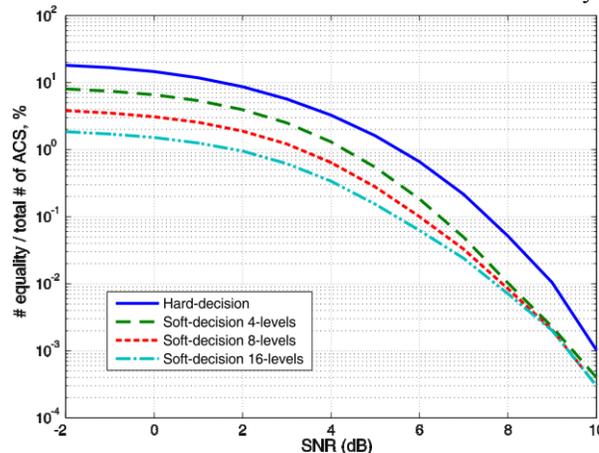

Fig. 3. The percentage of metric equality cases in total number of "compare" operations (PoZ = 0.5).

Toolbox [6] was used as the zero-oriented Viterbi decoder. BPSK-modulation along with AWGN channel has been used.

In order to reduce the simulation time the simple convolutional code with rate 1/2 and polynomials in octal form (5, 7) was considered. First question, which was investigated, is how often a decoder meets the metric equality cases. Fig. 3 shows numbers of the metric equality cases as a percentage of all "compare" operations for the hard-input decoder and the soft-input decoders with 4, 8 and 16 quantization levels. To obtain the soft-inputs uniform quantization was used. Fig. 3 shows stable correlation between the numbers of metrics equality cases and SNR level for all types of decoders. Obviously, the hard-input decoder has the maximum percentage of metric equality cases. Number of the metric equality cases also increases for low SNR values. Furthermore, it was found that PoZ does not affect the percentage of metric equality cases. Note, that PoZ = 0 means that the system encodes and decodes the message completely consisted of ones. Oppositely, PoZ = 1 represents all-zeros message.

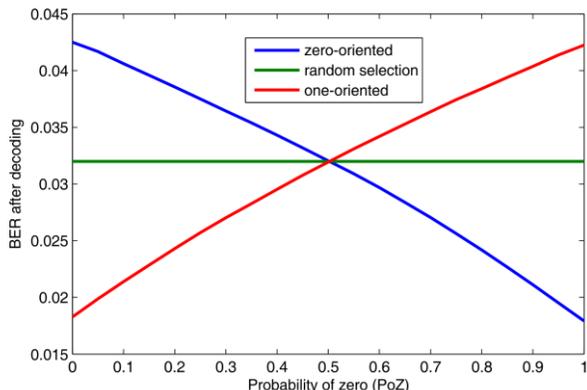

Fig. 4. Hard-input decoding performance against PoZ (SNR = 0 dB).

The impact of PoZ on the decoding performance for different implementations of VA can be treated as an important performance metric in the scope of identified effect. To access the influence hard-input decoder (Fig. 4) and soft-input decoders (Fig. 5) were simulated. For the soft-input decoder 8-levels uniform quantization was used, since it was shown that practically it is enough for optimum decoding [17]. Both figures were obtained under the same value of SNR=0 dB. Such low value was chosen for illustrative purposes: in order to clearly demonstrate the difference between the *zero-oriented* and the *one-oriented* approaches.

To reduce the simulation time the smooth filtering of the results was performed on the soft-input decoders (solid lines in Fig. 5). Both figures clearly show that the original VA with random path selection has the stable performance independently from the probability of zeros in input messages. However, the strict rule based Viterbi decoders show quite different behavior. Obviously, when only ones are transmitted (PoZ = 0), the one-oriented approach has the best performance while the zero-oriented approach has the worst performance. On the other hand, when only zeros are transmitted (PoZ = 1), the decoding performance of the zero-oriented approach is almost two times better than the performance of the original VA and 2.4 times better than the one-oriented decoder. It is important to note that for PoZ = 0.5 all three approaches show the same performance as expected.

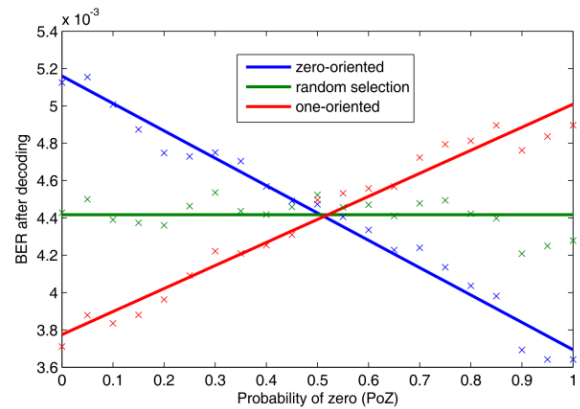

Fig. 5. Soft-input 8-levels decoding performance against PoZ (SNR = 0 dB).

Fig. 6 presents the decoding performance of different approaches for PoZ = 0 against SNR values ranging from -2 dB to 5 dB. The results demonstrate the maximum possible positive and negative effects with respect to the random selection. They are higher when the hard-input decoder is considered. This is explained by the fact that the chance of the metric equality for the hard-input decoder is higher (Fig. 3).

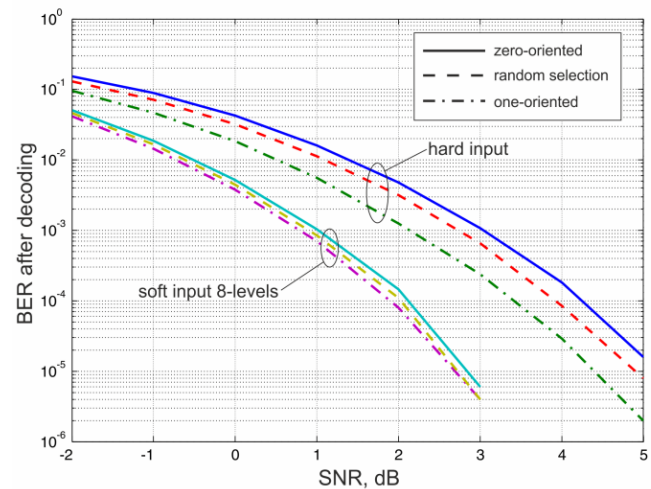

Fig. 6. Decoding performance against SNR (PoZ = 0).

Fig. 7 presents surfaces of the decoding performance for the zero-oriented Simulink Viterbi decoder. The hard-input decoder along with binary symmetric channel was used. The channel is characterized by the bit error probability. Five different convolutional codes were estimated. The decoding performance was assessed against the bit error probability and probability of zeros in input messages. The results confirm the observation that the decoding performance of the zero-oriented decoder depends on the input messages.

## V. DISCUSSION AND CONCLUSION

In this letter we reported the dependency of the decoding performance of Viterbi algorithm from characteristics of the input bit message. This dependency is a feature of VA implementation in the modern tools for experimental research. The substitution of the randomized tie-breaking step in the case of the paths' metric equality with strict logical conditions results in favorable decoding opportunities for messages with specific proportion of zeros and ones. This dependency was not documented previously in either the research literature or technical documentation. The roots for this implementation most likely go to early hardware implementations of the Viterbi algorithm [10] and an attempt of the experimental software to have as close to the reality implementation as possible. In the case of the hardware implementation, the choice of exchanging the randomization of the path selection with the strict logical condition is explained by substantial complications with designing electronic circuits for randomization. Although due to their proprietary designs it is hard to say whether commercial implementations feature the same VA implementation. However, Based on available sources [10,11,12,13,14,15,16] at least a part of modern hardware implementations is subject to the described dependency. This letter also identified that the open source SDR implementation – GNU radio [7] does so. Thus, the results reported in this paper could be practically used when simulating different VA decoders' implementations.

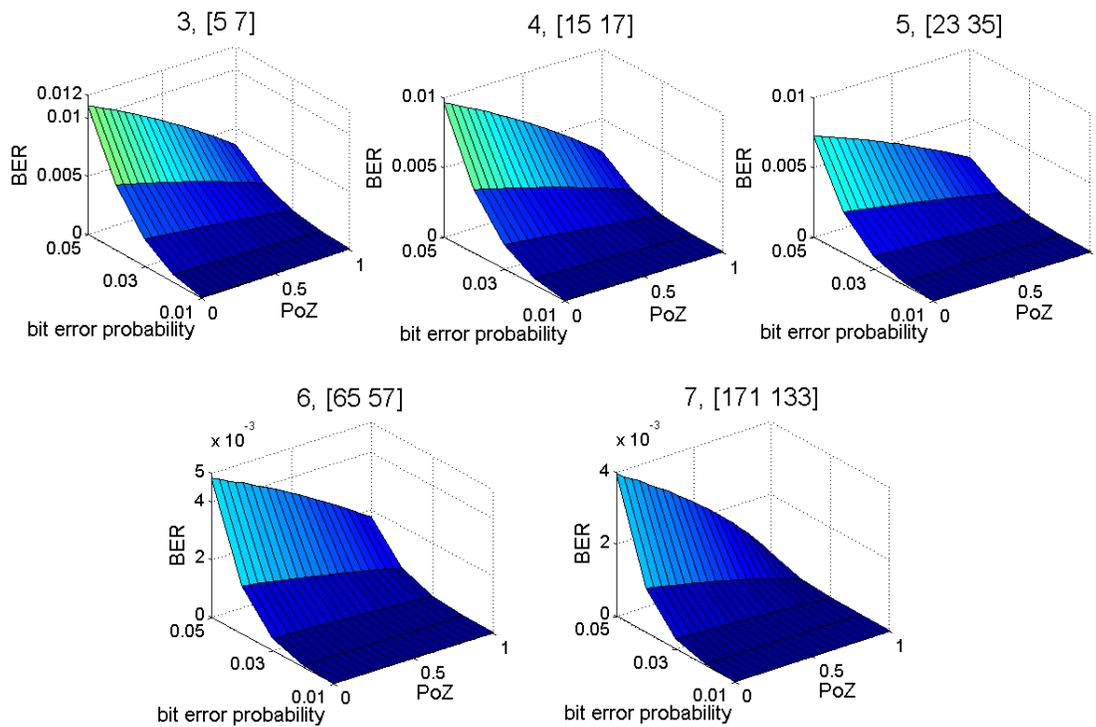

Fig. 7. The decoding performance of the built-in Simulink decoder for different polynomials.

To which extend the discovered dependency is critical in reality requires additional investigation. The performed simulations demonstrated the performance degradation in the considered scenarios could be severe. As a part of the discussion, however, it worth mentioning that several possible ways for improving the performance might be considered. For example, depending on the choice of the logical condition in the implementation one could deploy additional zero- or one-scrambling increasing by this the proportion of bits favorable by the implementation. Also the correlation between the count of the metric equality cases and SNR can be used to improve noise level estimation. Thus the number of observed metric equality cases could be converted into the SNR level. These topics as well as other possible optimization steps are subject for further investigations.